\documentclass[11pt,a4paper]{article}
 \usepackage{jheppub}

\usepackage{subfigure}
\usepackage{multirow}

\newcommand{\eqn}[1]{eq.~(\ref{#1})}
\newcommand{\Eqn}[1]{Eq.~(\ref{#1})}

\newcommand{\eqns}[2]{eqs.~(\ref{#1})-(\ref{#2})}

\newcommand{\YM}{Y_M}

\newcommand{\Ynu}{Y_\nu}

 \newcommand{\beq}{\begin{equation}}
 \newcommand{\eeq}{\end{equation}}
 \newcommand{\be}{\begin{equation}}
 \newcommand{\ee}{\end{equation}}
 \newcommand{\beqa}{\begin{eqnarray}}
 \newcommand{\eeqa}{\end{eqnarray}}
 \newcommand{\bea}{\begin{eqnarray}}
 \newcommand{\eea}{\end{eqnarray}}
 \newcommand{\no}{\nonumber}

 \newcommand{\cL}{\mathcal{L}}
 \newcommand{\cG}{\mathcal{G}}

\newcommand{\cO}{\mathcal{O}}

 \def\lsim{\ \rlap{\raise 3pt \hbox{$<$}}{\lower 3pt \hbox{$\sim$}}\ }
 \def\gsim{\ \rlap{\raise 3pt \hbox{$>$}}{\lower 3pt \hbox{$\sim$}}\ }
 \def\nnu{\nonumber}
 
\usepackage{color}

 \centerline{FTUAM-11-42 \qquad \qquad  IFT-UAM/CSIC-11-16 
\qquad \qquad TUM-HEP-797/11}

 \bigskip

 \title{Minimal flavour violation extensions of the seesaw}

 \author[a,b]{Rodrigo Alonso,} 
 \author[c]{Gino Isidori,} 
 \author[d,e]{Luca Merlo,} 
 \author[f,g]{Luis Alfredo Mu\~noz,} 
 \author[a,b,c]{and Enrico Nardi}

\affiliation[a]{Departamento de F\'{\i}sica Te\'orica,  
                C-XI, Facultad de Ciencias,  \\
                Universidad Aut\'onoma de Madrid, 
                C.U. Cantoblanco, 28049 Madrid, Spain}
\affiliation[b]{Instituto de F\'{\i}sica Te\'orica, IFT-UAM/CSIC \\
                Nicolas Cabrera 15, C.U. Cantoblanco, 28049 Madrid, Spain}
\affiliation[c]{INFN, Laboratori Nazionali di Frascati, \\
                Via Enrico Fermi 40,    I-00044 Frascati, Italy}
\affiliation[d]{{Physik-Department, Technische Universit\"at M\"unchen, \\
  	       James-Franck-Strasse, D-85748 Garching, Germany}}
\affiliation[e]{{TUM Institute for Advanced Study, Technische Universit\"at M\"unchen,\\
  	       Lichtenbergstrasse 2a, D-85748 Garching, Germany}}
\affiliation[f]{Instituto~de~F\'{i}sica,~Universidad~de~Antioquia\\  
                A.A. 1226 Medell\'\i n,~Colombia}
\affiliation[g]{ITM, Calle 73 No. 76A-354 Medell\'{\i}n, Colombia \\ }

\emailAdd{rod.alonso@estudiantes.uam.es}
\emailAdd{gino.isidori@lnf.infn.it}
\emailAdd{luca.merlo@ph.tum.de}
\emailAdd{lmunoz@fisica.udea.edu.co}
\emailAdd{enrico.nardi@lnf.infn.it}

\abstract{
We analyze the most natural formulations of the minimal
  lepton flavour violation hypothesis compatible with a type-I seesaw
  structure with three heavy singlet neutrinos $N$, and satisfying the
  requirement of being predictive, in the sense that all LFV effects
  can be expressed in terms of low energy observables.  We find a
  new interesting realization based on the flavour group
  $SU(3)_e\times SU(3)_{\ell+N}$ (being $e$ and $\ell$ respectively
  the $SU(2)$ singlet and doublet leptons). An intriguing feature of
  this realization is that, in the normal hierarchy scenario for
  neutrino masses, it allows for sizeable enhancements of $\mu \to e$
  transitions with respect to LFV processes involving the $\tau$
  lepton. We also discuss how the symmetries of the type-I seesaw
  allow for a strong suppression of the $N$ mass scale with respect to
  the scale of lepton number breaking, without implying a similar
  suppression for possible mechanisms of $N$ production.}


\begin{document} 
\maketitle
 \flushbottom

\section{Introduction}
\label{sec:intro}

The Minimal Flavour Violation (MFV) hypothesis~\cite{MFV,MFV2,DAmbrosio:2002ex} 
is, in short, the assumption that the sources of flavour symmetry breaking present in
the SM Lagrangian determine completely the structure of flavour
symmetry breaking also beyond the SM.  In the quark sector there is a
unique way to implement this hypothesis: the two quark SM Yukawa
couplings are identified as the only relevant breaking terms of the
$SU(3)^3$ quark-flavour symmetry~\cite{DAmbrosio:2002ex}.
  In contrast, in the lepton sector
there is no unique way to realize MFV.  The SM by itself cannot
accommodate Lepton Flavour Violation (LFV): since there is a single
set of Yukawa couplings (the ones for the charged leptons), they can
always be brought in diagonal form by rotating the three
$SU(2)_L$-doublets ($\ell_\alpha$) and the three right-handed (RH)
$SU(2)_L$-singlets $e_\alpha$ ($\alpha=e,\,\mu,\,\tau$).  However, LFV
is observed in neutrino oscillation experiments.  It is then
interesting to formulate an extension of the MFV hypothesis to the
lepton sector, or a Minimal Lepton Flavour Violation (MLFV)
hypothesis, whose starting point is not the SM Lagrangian but an
effective Lagrangian able to describe also the observed LFV effects in
the neutrino sector. The problem is that we do not know which physics
beyond the SM is responsible for the observed neutrino masses and LFV
effects, and different extensions of the SM correspond to different
formulations of the MLFV hypothesis.

The simplest way to extend the SM to include (strongly suppressed)
neutrino masses is by adding to the SM Lagrangian the 
dimension-five Weinberg operator~\cite{Weinberg:1979sa}: 
\bea
 \cL^{(m_\nu)}_{eff} &=&   \cL_Y^{\rm SM} + \cL_{D5}~,  \no \\
\cL_Y^{\rm SM} &=&
- \bar \ell_\alpha Y^{\alpha\beta}e_\beta\,  H   + {\rm h.c.},  \no \\
 \cL_{D5} &=& - \frac{g_{\alpha\beta}
}{2\,M}\left(\bar \ell^c_\alpha  \tilde H\right)
\left(\tilde H^T\ell_\beta\right) + {\rm h.c.},
  \label{eq:SMD5}
\eea
where $H$ is the Higgs field,\footnote{~Here and in the following the
  indexes on the lepton fields refer only to the flavour structure;
  $\tilde H = i\tau_2 H^*$, $\ell^c=C\gamma^0 \ell^*$, and the
  appropriate $SU(2)_L$ index contraction with the lepton doublets
  $\ell$ is left understood.}  and $M$ is a high scale related to the
breaking of the lepton number ($L$).  Although the appearance of
neutrino masses and LFV is linked to the introduction of a
non-renormalizable operator, the MLFV formulation based on
\Eqn{eq:SMD5} has a minimal field content and is minimal also in terms
of the relevant flavour symmetry. The latter can be chosen to be
$U(3)_\ell \times U(3)_e$: the largest symmetry group of the gauge
invariant kinetic terms of the SM leptons.  Factorizing the two $U(1)$
groups identified by lepton number and hypercharge, under the
remaining semi-simple flavour subgroup $SU(3)_\ell \times SU(3)_e$ the
leptons transform as $\ell \sim (3,1)$ and $e \sim (1,3)$.  Formal
invariance of the effective Lagrangian under $SU(3)_\ell \times
SU(3)_e$ is then recovered by promoting the couplings $Y_{\alpha\beta}$
and $g_{\alpha\beta}$ to spurion fields with the assignments $Y\sim
(3, \bar 3)$ and $g\sim (\bar6,1)$.

The LFV operators of dimension-six, that are naturally present in the
effective theory approach, conserve $B-L$~\cite{Weinberg:1979sa} and
are suppressed by a new effective scale $\Lambda$ not necessarily
related to $M$ that, in this minimal scheme, is the $L$-breaking
scale.  The extremely tight limits on $B$ violating processes then
imply that if $\Lambda\ll \Lambda_{GUT}$, then the dimension-six
operators must conserve $B$, and thus $L$ as well.  According to the
MLFV ansatz, these operators are built only in terms of SM fields and
spurions, preserving formal invariance under the flavour
group. Besides the requirement of a sufficiently low scale $\Lambda$,
the possibility of observing new LFV effects also requires rather
large values of $g_{\alpha\beta}$. Since the magnitude of
$g_{\alpha\beta}/M$ is fixed by the light neutrino mass scale, a large
$g_{\alpha\beta}$ requires a correspondingly high scale $M$, and this
results in a large hierarchy $\Lambda/M\ll 1$.  A detailed study of
this framework is given in~\cite{Cirigliano:2005ck,Cirigliano:2006su}
and will not be repeated here.

The main drawback of the MLFV ansatz based on~\Eqn{eq:SMD5} is that it
cannot be linked to several dynamical models for neutrino masses based
on the seesaw mechanism.  This is why in~\cite{Cirigliano:2005ck}, and
later on also in~\cite{Cirigliano:2006nu}, a second scenario, with an
extended field content and a different flavour group has been
analyzed. Alternative formulations of the MLFV hypothesis have also
been proposed in~\cite{Davidson:2006bd,Gavela:2009cd}.  Once the field
content of the theory is extended, there are in principle several
possibilities to define the flavour symmetry and a consistent minimal
set of spurions. However, if we restrict the attention to the popular
type-I seesaw models~\cite{seesaw}, the choice is restricted: the
purpose of this paper is to analyze the most natural formulations of
the MLFV hypothesis compatible with a type-I seesaw structure with
three singlet Majorana neutrinos. We find a new MLFV realization that
allows for sizable enhancements of the LFV violating processes
involving the lighter generations, and thus is phenomenologically
interesting.

\section{Minimal effective theories for the seesaw}

In order to define a MLFV effective theory in the context of a type-I
seesaw structure, we assume that in addition to the SM leptons ($\ell$
and $e$) at high energies there is at least another set of dynamical
fields carrying lepton flavour: the three SM singlets heavy Majorana
neutrinos $N$.  The largest group of flavour transformations commuting
with the gauge invariant kinetic terms of the lepton fields $\ell$,
$N$ and $e$ is $\cG = U(3)_\ell\times U(3)_N\times U(3)_e$.  We assume
that $\cG$, or some subgroup of $\cG$, is the relevant group of
flavour transformations, and that this symmetry is broken at some high
scale $\Lambda_F$ larger than the scale of the RH neutrino
masses. Most important, we require that all the relevant
symmetry-breaking terms (the spurions) can be identified with the
couplings appearing in the seesaw Lagrangian, namely the
renormalizable mass terms of the three basic sets of lepton fields.

Integrating out the heavy degrees of freedom, the effective Lagrangian
relevant at energies below $\Lambda_F$ can be decomposed as
\be
\label{eq:2}
\cL_{\rm eff} (E < \Lambda_F) = \cL_{\rm kin}(N,\ell,e) +
\cL_{\rm seesaw}(N,\ell,e;H)  + \Delta \cL_{\Lambda_F}~,
\ee
where
\be
- \cL_{\rm seesaw}(N,l,e;H) = 
\epsilon_e\, \bar \ell_\alpha Y_e^{\alpha\beta}e_\beta\,  H   
+ \epsilon_\nu\, \bar\ell_\alpha Y_\nu^{\alpha j}\,N_j\, \tilde H
+\frac{1}{2}\,\epsilon_\nu^2\,\mu_L\; \bar N^c_i\, Y_M^{ij}\, N_j+{\rm h.c.}.   
\label{eq:seesaw}
\ee
$\Delta \cL_{\Lambda_F}$ in \eqn{eq:2} denotes higher-dimensional
operators involving $N$, $\ell$, and $e$, suppressed by inverse powers
of ${\Lambda_F}$, as well as other renormalizable and
non-renormalizable interactions of these lepton fields with possible
additional degrees of freedom that are relevant below
$\Lambda_F$.
In order to analyze the transformation properties of the $\cG$-breaking 
spurions appearing in $\cL_{\rm seesaw}$, we decompose 
the symmetry group as follows
\be
\cG = U(1)_Y \times U(1)_L \times U(1)_R \times \cG_F~, \qquad 
\cG_F = SU(3)_\ell\times SU(3)_N\times SU(3)_e~,
\ee
where $U(1)_Y$ and $U(1)_L$ correspond to hypercharge (that remains
unbroken) and to total lepton number, respectively. The remaining
Abelian factor can be identified either with $U(1)_e$ or with
$U(1)_N$, corresponding respectively to global phase rotations of the
RH charged leptons or RH neutrinos, and we will generically denote it
by $U(1)_R$.  By construction, we assume that $Y_{e,\nu,M}$ are
dimensionless spurions that carry no $U(1)_R$ charges, and thus break
only $\cG_F$.  The transformation properties of the lepton fields and
of the spurions under $\cG_F$ are
\begin{eqnarray}
  \label{eq:fields}
\ell  \sim (3\,,1\,,1) 
\qquad& 
N \sim (1\,,3\,, 1) 
&\qquad 
e \sim (1\,,1\,,3)~,
\\ 
  \label{eq:generalspurions}
Y_\nu \sim (3\,,\bar 3\,,1)
\qquad&  
Y_M \sim (1\,,\bar 6\,,1) 
& \qquad 
Y_e \sim (3\,,1\,,\bar 3)\,. 
\end{eqnarray}

As regards the two broken Abelian factors, $U(1)_L$ is broken (by two
units) by $\mu_L$, that is a spurion with dimension of a mass, while
$U(1)_R\,$
is broken by a dimension-less spurion $\epsilon_R\,$, where
$\epsilon_R$ denotes $\epsilon_e$ or $\epsilon_\nu$.

At this point we are ready to make one more step, integrating out from
the effective Lagrangian~\eqn{eq:2} the heavy RH neutrinos with masses
of order $ \epsilon^2_\nu\mu_L \ll \Lambda_F$ and any other non-SM
field, that we assume with masses at some scale $\Lambda \ll
\Lambda_F$, with $\Lambda$ around or above the electroweak scale.  The
resulting effective Lagrangian can be decomposed as 
\be 
\cL_{\rm eff}
(E < \Lambda) = \cL_{\rm SM} + \cL^{\rm seesaw}_{D5} +
\frac{1}{\Lambda^2} \sum_{i} c_i O_i^{(6)} + \ldots~, 
\ee 
where
$\cL^{\rm seesaw}_{D5}$ is nothing but the Weinberg operator, whose
coupling can now be determined in terms of the spurions appearing in
\Eqn{eq:seesaw}.
The $O_i^{(6)}$ denote generic dimensions-six operators written in
terms of the SM fields and of the spurions, and the dots denotes
operators of higher dimension. As shown in~\cite{Weinberg:1979sa},
dimensions-six operators written in terms of the SM fields conserve
$B-L$, and since we have not introduced (dangerous) sources of $B$
violation, then the operators $O_i^{(6)}$ must  conserve separately
$L$.  This is the reason why the scale $\Lambda$ can be substantially
lower than $\Lambda_F$ and of the RH neutrino mass scale.

As far as the flavour structure of the $O_i^{(6)}$ is concerned, 
our assumptions about the breaking of $\cG_F$ imply the following
\begin{itemize}
\item[{\bf I.}] {\em All higher-dimensional operators must be formally
    invariant under $\cG_F$ once the transformation properties of the
    fields eq.}~(\ref{eq:fields}) {\it and of the spurions eq.}
  (\ref{eq:generalspurions})~{\it are taken into account}.
\end{itemize}
As is pointed out in~\cite{Cirigliano:2005ck}, this condition alone is
not sufficient to obtain an effective theory that is predictive, since
the flavour structure of the three spurions $Y_\nu,\,Y_M$ and $Y_e$
cannot be determined from low-energy data alone.  A predictive MLFV
formulation must satisfy an additional working hypothesis:
\begin{itemize}
\item[{\bf II.}]
 {\em The flavour structure of the spurions must be
  determined in terms of low energy observables, namely 
  the PMNS mixing matrix and the light neutrino mass eigenvalues.}
\end{itemize}
The only way this second hypothesis can be satisfied is by restricting
the form of the spurions $Y_i$ in such a way that the relevant LFV
combinations will depend on a reduced number of parameters.  As we show
in the following, this goal can be naturally obtained by assuming that
the underlying flavour symmetry corresponds to  a subgroup of  $\cG_F$ 
rather than to the full flavour group $SU(3)^3$.

\mathversion{bold}
\subsection{Breaking of the $U(1)$ symmetries and size of the LFV effects}
\mathversion{normal}

Before analyzing the possible subgroups of  $\cG_F$ leading to predictive 
frameworks, it is worth to discuss in general terms the overall size of the 
LFV effects and its connection to the breaking of  $U(1)_R$. 
The explicit structure of the Weinberg operator
obtained by integrating-out the RH neutrinos, and the corresponding
light neutrino mass matrix, are
\be
\label{eq:Mnu}
\cL^{\rm seesaw}_{D5} = \frac{1}{\mu_L}
\left(\bar\ell\tilde H\right) 
Y_\nu\frac{1}{Y_M}Y_\nu^T  \left(\tilde H^T\ell^c\right)  
\quad \longrightarrow  \quad
  m_\nu^\dagger = 
\frac{v^2}{\mu_L}\; Y_\nu\frac{1}{Y_M}Y_\nu^T
= U\, \mathbf{m}_\nu\, U^T    
\ee
where $v=\langle H\rangle$ is the Higgs vacuum expectation value (vev), $U$ is the PMNS matrix
and ${\mathbf m}_\nu = {\rm diag }
(m_{\nu_1},\,m_{\nu_2}\,,m_{\nu_3})$.  Note that since the Weinberg
operator does not break $U(1)_{R}$, the overall size of ${\mathbf
  m}_\nu$ depends only on the lepton-number violating scale $\mu_L$,
but not on $\epsilon_{e,\nu}$. 
Without loss of generality we can rotate $Y_{e}$ and $Y_M$ 
to a basis where they are both diagonal, and in terms of mass 
eigenvalues they can be written as:
\begin{eqnarray}
  \label{eq:diage}
Y_e&=&\frac{1}{\epsilon_e\,v}\,{\rm  diag}(m_e,m_\mu,m_\tau)\,, \\ 
  \label{eq:diagM}
Y_M&=&\frac{1}{\epsilon_\nu^2\,\mu_L}\,{\rm  diag}(M_1,M_2,M_3)\,.  
\end{eqnarray}
These equations show that the overall size of $Y_e$ and $Y_M$ is
controlled by the Abelian spurion, and the same is true for $Y_\nu$.
A natural choice for the size of the $U(1)_R$ breaking is the one that
allows us to keep $\cO(1)$ entries in the $Y_i$ matrices. In the case
of the light-neutrino mass matrix, this choice unambiguously points to
a very large $L$-breaking scale
\be
\label{eq:muL}
\mu_L \sim \frac{v^2}{\sqrt{\Delta m^2_{\rm atm}}} \approx  6\times 10^{14}~{\rm GeV}.
\ee
As far as $\epsilon_e$ and $\epsilon_\nu$ are concerned, we can
envisage two possibilities, depending if the additional broken Abelian
symmetry is $U(1)_e$ or $U(1)_N$.  In the first case we can set
$\epsilon_e \approx m_\tau/v$, providing a natural explanation for the
smallness of the charged-lepton Yukawa coupling, but then naturalness
suggests $\epsilon_\nu \approx 1$, i.e.~very heavy RH neutrinos with
masses of $\cO(\mu_L)$.  In the second case $\epsilon_e \approx
1$,\footnote{~If $\epsilon_e \approx 1$, the suppression of the
  charged-lepton masses could still be justified in a multi-Higgs
  scenario by the hierarchy in the vevs with different hypercharge:
  $v_d \ll v_u$.}  but we are free to assume $\epsilon_\nu \ll 1$ as
would naturally result from an approximate $U(1)_N$ symmetry. In this
case the RH neutrinos could have masses well below the $L$-breaking
scale, and possibly within the reach of future experiments.  From the
phenomenological point of view this second one is clearly the most
interesting choice, and is the one we will adopt from now on.

As we have already discussed, the dimension-six operators contributing
to low-energy LFV processes are invariant under $U(1)_L$ and, by
construction, they are also invariant under $U(1)_N$. As a result,
assuming that the $Y_i$ spurions have $\cO(1)$ entries implies that
the overall scale of LFV effects is controlled only by the effective
scale $\Lambda$.  The generic structure of the most relevant LFV
operators is
\bea O^{(6)}_{LR} &=& \bar \ell_\alpha
(\Delta_{8} Y_e)^{\alpha\beta} (\sigma\cdot F) e_\beta~H~,
\label{eq:O1}
\\
O^{(6)}_{LL1} &=& \bar \ell_\alpha \Gamma \Delta_{8}^{\alpha\beta} \ell_\beta \times  \bar f  \Gamma^\prime f  
\label{eq:O2}
\\
O^{(6)}_{LL2} &=& \bar \ell_\alpha \Gamma \Delta_{8}^{\alpha\beta} \ell_\beta \times 
\bar \ell_\alpha \Gamma^\prime \Delta_{8}^{\alpha\beta} \ell_\beta~,
\label{eq:O3}
\\
O^{(6)}_{LL3} &=& \bar \ell_\alpha \Gamma \Delta_{6}^{\alpha\beta} \ell^c \times
\bar \ell^c_\alpha \Gamma^\prime \Delta_{\bar 6}^{\alpha\beta} \ell_\beta~,
\label{eq:O4}
\eea
where $\Delta_{8}$, $\Delta_{6}$, and $\Delta_{\bar 6}$ are $SU(3)_e
\times SU(3)_N$ singlets combinations of the $Y_i$ transforming as
$8$, $6$ and $\bar 6$ of $SU(3)_\ell$, respectively ($\Gamma$ stands
for generic Dirac structures and/or $SU(2)_L$ matrices, $F$
generically denotes the field strength of $SU(2)_L \times U(1)_Y$
gauge fields, while $f$ stands for SM leptons or quarks). Because of
the hierarchical structure of $Y_e$, LFV bilinears with two RH charged
leptons (such as $\bar e Y_e \Delta_{8} Y_e e$) are always suppressed
with respect to the corresponding LH terms in
Eqs.~(\ref{eq:O2})--(\ref{eq:O4}) and we neglect them.

Considering terms with up to two $Y_{\nu}$ and two $Y_M$, we can write
the following contributions to the operators \eqns{eq:O1}{eq:O4}:
\be 
\Delta^{(1)}_{8} = \Ynu\Ynu^\dagger~,~ 
\qquad\quad\  
\Delta^{(1)}_{6} = \Ynu\YM^\dagger \Ynu^T~,~ 
\qquad\quad\  
\Delta^{(2)}_{8} ~=~ \Ynu \YM^\dagger\YM \Ynu^\dagger~,  
\ee
with $\Delta^{(1)}_{\bar 6}= (\Delta^{(1)}_{6})^\dagger$.
In the mass diagonal basis of \eqns{eq:diage}{eq:diagM} all LFV
effects are associated to $Y_\nu$, which is a generic complex $3\times
3$ matrix corresponding to 15 physical parameters. In the absence of
further assumptions we will not be able to determine all these
parameters from \Eqn{eq:Mnu}.  However, as anticipated, predictive
frameworks can be obtained by choosing as the underlying flavour
symmetry some suitable subgroup of $\cG_F$.

\subsection{Two predictive cases}

There are basically two natural criteria that we can follow 
to relate the LFV structures $\Delta$ to the observables 
in \Eqn{eq:Mnu}. The two criteria, which can be formulated in terms 
of general symmetry hypotheses, allow us to assume 
that in a given basis either $Y_\nu$ or $Y_M$ corresponds to  
the identity matrix in flavour space $I_{3\times 3}$.
\begin{itemize}
\item[{\bf A.}] {\it $SU(3)_\ell\times  SU(3)_N \to SU(3)_{\ell+N}$.}\\ [5pt]
  If we assume that $\ell$ and $N$ belongs to the fundamental
  representation of the same $SU(3)$ group, then in a generic basis
  $Y_\nu$ must be a unitary matrix (and thus it can be always rotated
  to the identity matrix by a suitable unitary transformation of the
  RH neutrinos).  This condition  is sufficient to allow inverting the seesaw
  formula in \eqn{eq:Mnu}.  By doing so we find
\begin{eqnarray}
  \label{eq:MMLFV1}
\Delta_6 &=&  \Delta^\dagger_{\bar 6} = 
\left[\left(Y_\nu\frac{1}{Y_M}Y_\nu^T\right)^{-1}\right]^{\dagger}
  =\frac{v^2}{\mu_L} \; 
  U\, \frac{1}{\mathbf{m}_\nu}\, U^T \,, \\
  \label{eq:MMLFV2}
\Delta_8^{(2)} &=&  \Delta_6 \cdot \Delta_6^\dagger
  =\frac{v^4}{\mu_L^2} \; 
  U\, \frac{1}{\mathbf{m}_\nu^2}\, U^\dagger~,
\end{eqnarray}
while $\Delta_8^{(1)} =I_{3\times 3}$ and gives no LFV effects.

The choice of a unitary $Y_\nu$ can also be justified on a different
basis.  According to a general theorem~\cite{Bertuzzo:2009im} if the
$N$'s belong to an irreducible 3-dimensional representation of a
non-Abelian group, then $Y_\nu$ is (proportional to) a unitary matrix.
Let us recall that models for neutrino masses based on discrete
non-Abelian flavour symmetries have proved to be quite successful in
reproducing the structure of the PMNS matrix.  This is generally
related to the fact that in first approximation the symmetry implies a
tri-bimaximal (TBM)~\cite{TBM} mixing pattern that is a good
approximation to PMNS.  We can then picture a situation where in a
first step of the flavour symmetry breaking $SU(3)_N$ breaks to a
non-Abelian discrete subgroup having irreducible 3-dimensional
representations to which the $N$'s are assigned. In this case $Y_\nu$
can be non-trivial but must be proportional to a unitary matrix (while
$Y_M$ is clearly $\propto I$ or vanishing).  In a second step, when
the discrete symmetry is broken, $Y_M$ acquires a non-trivial
structure, while corrections to $Y_\nu$ can be quantified to remain at
the level of the deviations of $U$ from TBM, that is small.  Several
models based on discrete non-Abelian symmetry that yield a unitary
$Y_\nu$ or $Y_\nu \propto I_{3\times 3}$ have been constructed, and a
long list of references, properly classified according to these two
possibilities, can be found in~\cite{AristizabalSierra:2009ex}.

A detailed phenomenological analysis of this scenario is presented in
the next session. The main distinctive feature with respect to the
case based on the $O(3)_N$ symmetry analyzed
in~\cite{Cirigliano:2005ck} (see point B.~below) is that, due to the
inverse $\mathbf{m}_\nu$ dependence in
Eqs.~(\ref{eq:MMLFV1})--(\ref{eq:MMLFV2}), LFV processes are {\it
  enhanced} when the lighter neutrinos mass eigenvalues are
involved. This implies, in particular, a potentially strong
enhancement of $\mu\to e\gamma$ in the normal-hierarchy (NH) case.

Other phenomenologically interesting features of this scenario, that
are largely independent of the particular pattern of flavour symmetry
reduction $SU(3)_\ell\times SU(3)_N \to SU(3)_{\ell+N} $ but are
mainly related to the breaking of $U(1)_N$ and to assumptions about
the size of the spurion $\epsilon_N$, are:
\begin{itemize}

\item Being unrelated to the breaking of $U(1)_L$ and $U(1)_N$, the
  LFV violating scale $\Lambda$ can be as low as permitted by present
  phenomenological constraints.

\item Assuming ${\cal O}(1)$ entries for the non-Abelian spurions
  $Y_i$, double LFV processes (such as $\tau \to e e \bar \mu$) are
  not necessarily strongly suppressed with respect to single LFV
  process (such as $\tau \to e \bar e \mu $).

\item As long as $\epsilon_\nu \gg v/\mu_L$ the light neutrino masses
  do not depend on its value. In contrast, the masses of the RH
  neutrinos are suppressed with respect to $\mu_L$ by two powers of
  $\epsilon_\nu$. Then, even if the $L$-number breaking scale $\mu_L$
  is generically large~\eqn{eq:muL}, if the $U(1)_N$ breaking is small
  ($\epsilon_\nu \lsim 10^{-5}$)  $N$ states with masses of a few
  TeV (or even lower) are an open possibility.

\item The symmetries of this scenario imply that $U(1)_N$ conserving
  operators of the form $(\bar N\, N)\cdot(\bar q\, q)$ (where $q$
  denote quark fields) are not suppressed by any power of $\epsilon_\nu$.
  Therefore, in the absence of other  suppressing effects, we can even
  envisage the possibility that the $N$'s can be produced at
  colliders.

\end{itemize}

\item[{\bf B.}] {\it  $SU(3)_N \to O(3)_N \times CP$.} \\ [5pt]
  Assuming that the flavour group acting on the RH neutrinos is
  $O(3)_N$ rather than $SU(3)_N$, implies that $Y_M$ must be
  proportional to $I_{3\times 3}$. However, this condition alone is
  not enough to deduce the structure of $Y_\nu$ from the seesaw
  formula: this requirement, (and hence the predictivity of the
  theory) is fulfilled only if we further assume that $Y_\nu$ is real
  $Y_\nu^\dagger = Y_\nu^T$ (which follows from imposing CP
  invariance)~\cite{Cirigliano:2005ck}. In this case, since the
  Majorana mass term has a trivial structure, all LFV effects stem
  from the (real) Yukawa coupling matrices:
\begin{equation}
  \label{eq:ON}
  \Delta_6 = \Delta_8^{(1)}=    \Delta_8^{(2)}= 
Y_\nu Y_\nu^T= \frac{\mu_L}{v^2}\; 
 U\, \mathbf{m}_\nu\, U^T \,. 
\end{equation}
The implications for LFV processes of this scenario have been analyzed 
in~\cite{Cirigliano:2005ck} and we refer to this paper for further details.

\end{itemize}

\subsection{Phenomenology}

\begin{figure}[t!]
  \centering
  \includegraphics[width=7.5cm,height=6.0cm]{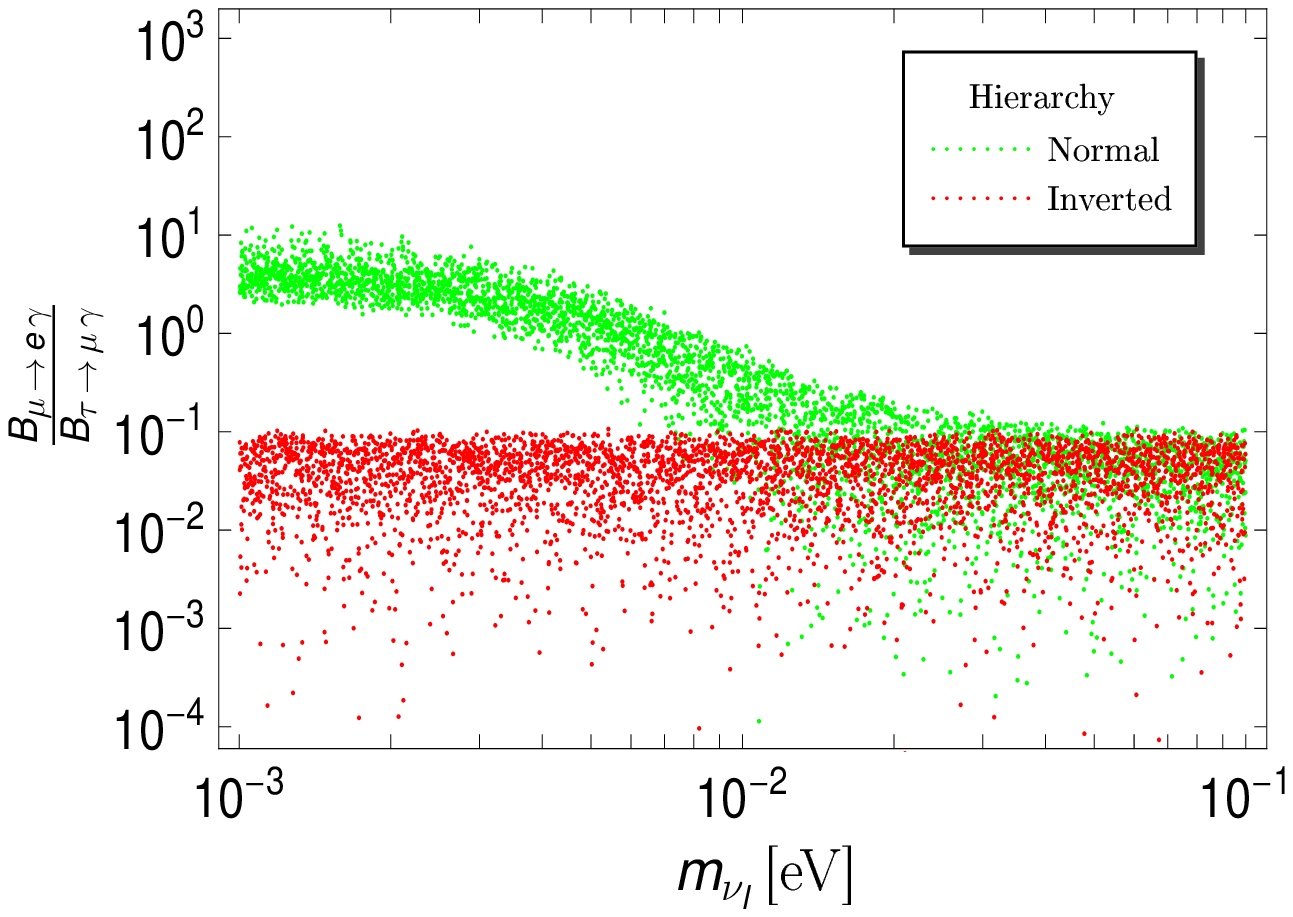}
  \includegraphics[width=7.5cm,height=6.0cm]{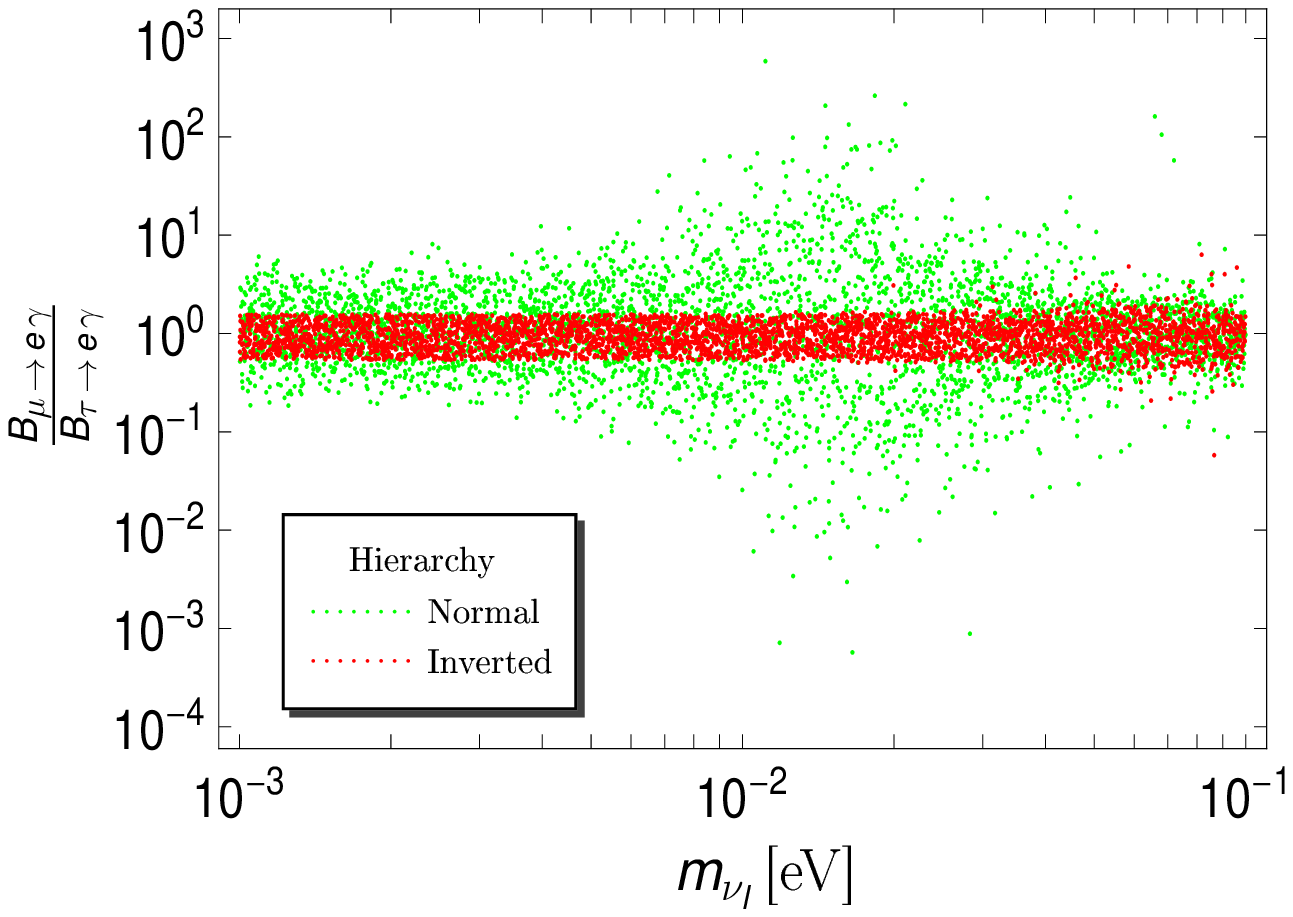}
  \caption{Scatter plots of the ratios
    $\frac{B_{\mu\,\rightarrow\,e\,\gamma}}{B_{\tau\,\rightarrow\,\mu\,\gamma}}$
    (left) and
    $\frac{B_{\mu\,\rightarrow\,e\,\gamma}}{B_{\tau\,\rightarrow\,e\,\gamma}}$
    (right) as a function of the mass of the lightest neutrino.  Green
    points correspond to NH with $m_{\nu_l}=m_{\nu_1}$.  Red points
    correspond to IH with $m_{\nu_l}=m_{\nu_3}$.  The density of
    points depends on arbitrary details of the sampling procedure
    and does not represent the likelihood of different regions.}
\label{fig::brm1}  
\end{figure}
In this section we discuss the dependence of LFV processes, and in
particular of the radiative decay $\ell_i\to \ell_j\,\gamma$, on
$\Delta^{(2)}_8$ defined in \eqn{eq:MMLFV2}.  In order to
compare the relevance of different decay channels we define the
normalized branching fractions~\cite{Cirigliano:2005ck}:
\begin{equation}
\label{eq:Br}
 B_{\ell_i\to \ell_j\gamma} \equiv  
\frac{\Gamma(\ell_i\to \ell_j\gamma)}{\Gamma(\ell_i\to \ell_j\nu_i\bar{\nu}_j)}.
\end{equation}
We are interested in studying quantitatively  ratios of these quantities 
for different types of radiative decays. These ratios  simply reduce to  
ratios of the modulus squared  of the corresponding $\Delta^{(2)}_8$ entries: 
\begin{equation}
\frac{B_{\ell_i\,\rightarrow\,\ell_j\,\gamma}}
{B_{\ell_k\,\rightarrow\,\ell_m\,\gamma}}=
\frac{\left|\left( \Delta^{(2)}_8\right)_{ij}\right|^2}
{\left|\left(\Delta^{(2)}_8\right)_{km}\right|^2}\,, 
\end{equation}
Omitting the prefactor $\frac{v^4}{\mu_L^2}$ that cancels in the
ratios, the relevant LFV structures then reduce to $\Delta_8^{(2)} \to
U\,\frac{1}{\mathbf{m}_\nu^2}\,U^\dagger$.  We generate random values
for these quantities allowing the neutrino parameters to vary within
their (approximate) 2$\sigma$ c.l. experimental intervals~\cite{nudata}:
\begin{eqnarray}
   \label{eq:data} 
\nnu
\Delta\,m^2_{\rm sol}:&&(7.3\,-\,8.0)\times10^{-5}\,{\rm eV}^2\,, \\ \nnu
\Delta\,m^2_{\rm atm}:&&(2.2\,-\,2.6)\times10^{-3}\,{\rm eV}^2\,, \\ \nnu
\sin^2\theta_{12}:&&     0.28\,-\,0.35       \,,      \\ \nnu
\sin^2\theta_{23}:&&        0.35\,-\,0.61    \,,        \\ 
\sin^2\theta_{13}:&&        0.0   \,-\ 0.04\,.       
\end{eqnarray}

For the NH and inverted hierarchy (IH) we restrict the range of
variation of the lightest mass eigenvalue respectively to $
m_{\nu_1}\, (m_{\nu_3}) \leq 0.1\,$eV, while the CP phase $\delta$,
that enters all the formulas through $\cos\delta$,  is
varied in the interval $[0\,-\,\pi]$ (the Majorana phases
are of course irrelevant for  $\Delta L=0$ processes).

In Fig.~\ref{fig::brm1} we present the results for the ratios
$\frac{B_{\mu\,\rightarrow\,e\,\gamma}}{B_{\tau\,\rightarrow\,\mu\,\gamma}}$
(left panel) and
$\frac{B_{\mu\,\rightarrow\,e\,\gamma}}{B_{\tau\,\rightarrow\,e\,\gamma}}$
(right panel) as a function of the lightest mass eigenvalue
$m_{\nu_l}=m_{\nu_1}$ (NH: green points) and $m_{\nu_l}=m_{\nu_3}$
(IH: red points) while all the other parameters are varied aleatorily
in the given intervals.  Note that
 in this figure, as well as in all
the other figures below, the density of points depends on arbitrary
details of the sampling procedure, and should not be interpreted as
related to the the likelihood of regions differently populated.

From the first panel we see that for NH and small values of $m_{\nu_1}
\lsim 10^{-2}\,$eV we generically have
$B_{\mu\,\rightarrow\,e\,\gamma}\,>\,B_{\tau\,\rightarrow\,\mu\,\gamma}$.
The enhancement of $B_{\mu\,\rightarrow\,e\,\gamma}$ is obviously due
to ${{\bf m}^2_\nu}$ appearing in the denominator of $\Delta^{(2)}_8$,
and can be of a factor of a few. In the limit of $m_{\nu_1}\ll
m_{\nu_{2,3}}$ we have:
\begin{equation}
\label{B1}
\frac{B_{\mu\,\rightarrow\,e\,\gamma}}{B_{\tau\,\rightarrow\,\mu\,\gamma}}
\ \sim\  
\frac{c^2_{12}c^2_{13}}{\left(c_{12} c_{23}s_{13}\mp s_{12}s_{23}\right)^2}
\ \approx \  7.3\; (3.2) \,, 
\end{equation} 
where $c_{ij}\equiv \cos\theta_{ij}$ and $s_{ij}\equiv
\sin\theta_{ij}$ and the $-$ ($+$) sign in the denominator of the
first equality corresponds to $\delta=0$ ($\delta=\pi$). The numerical
estimate in the last equality is obtained using the best fit values of
the mixing angles, again for $\delta=0$ ($\delta=\pi$).  When
$m^2_{\nu_1}\gg \Delta\,m^2_{\rm sol}$ and $m_{\nu_1} \approx
m_{\nu_2}$, the contributions to $\mu\,\rightarrow\,e\,\gamma$
proportional to $\theta_{12}$ suffer a strong GIM suppression, and the
decay rate becomes proportional to $\theta_{13}^2$ .  This behavior
is seen clearly in Fig.~\ref{fig::brm1} (left) for values of
$m_{\nu_1} \approx 10^{-2}\,$eV.

For the IH, in the limit $m_{\nu_3}\ll m_{\nu_{1,2}}$ and
independently of the value of $\delta$ we obtain: 
\begin{equation}
\label{eq:B1a}
\frac{B_{\mu\,\rightarrow\,e\,\gamma}}{B_{\tau\,\rightarrow\,\mu\,\gamma}}
\ \sim\ 
\frac{s^2_{13}}{c^2_{13}c^2_{23}}
\ \approx\  2\, s_{13}^2 \,.
\end{equation} 
Approximately the same  result is obtained also 
in the limit of large masses $m_{\nu_i} \gg \sqrt{\Delta m^2_{\rm atm}}$,  
which explains why for $m_{\nu_1}\to 10^{-1}\,$eV the results for  IH and NH 
converge. 
\begin{figure}[t!]
  \centering
  \includegraphics[width=7.5cm,height=6.0cm]{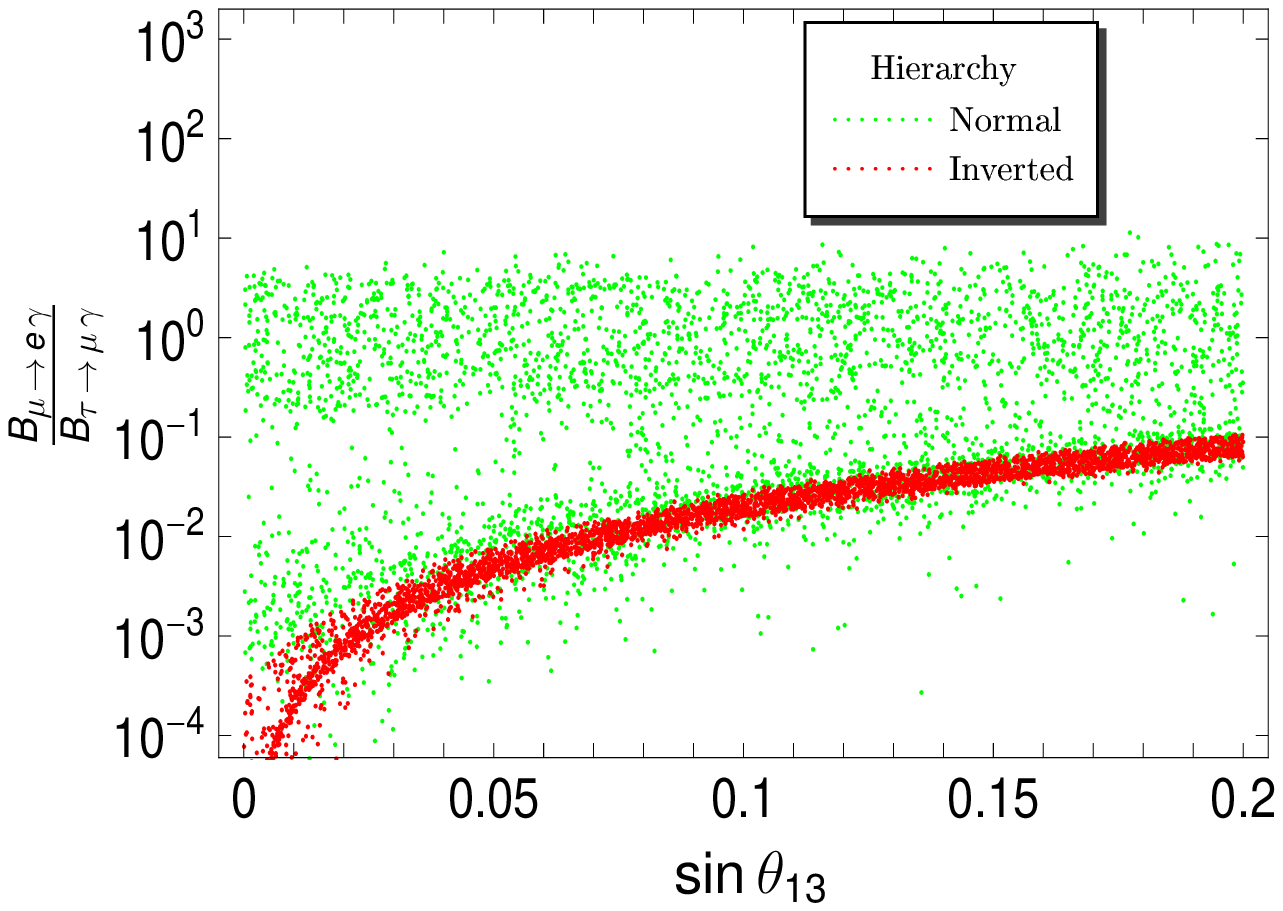}
  \includegraphics[width=7.5cm,height=6.0cm]{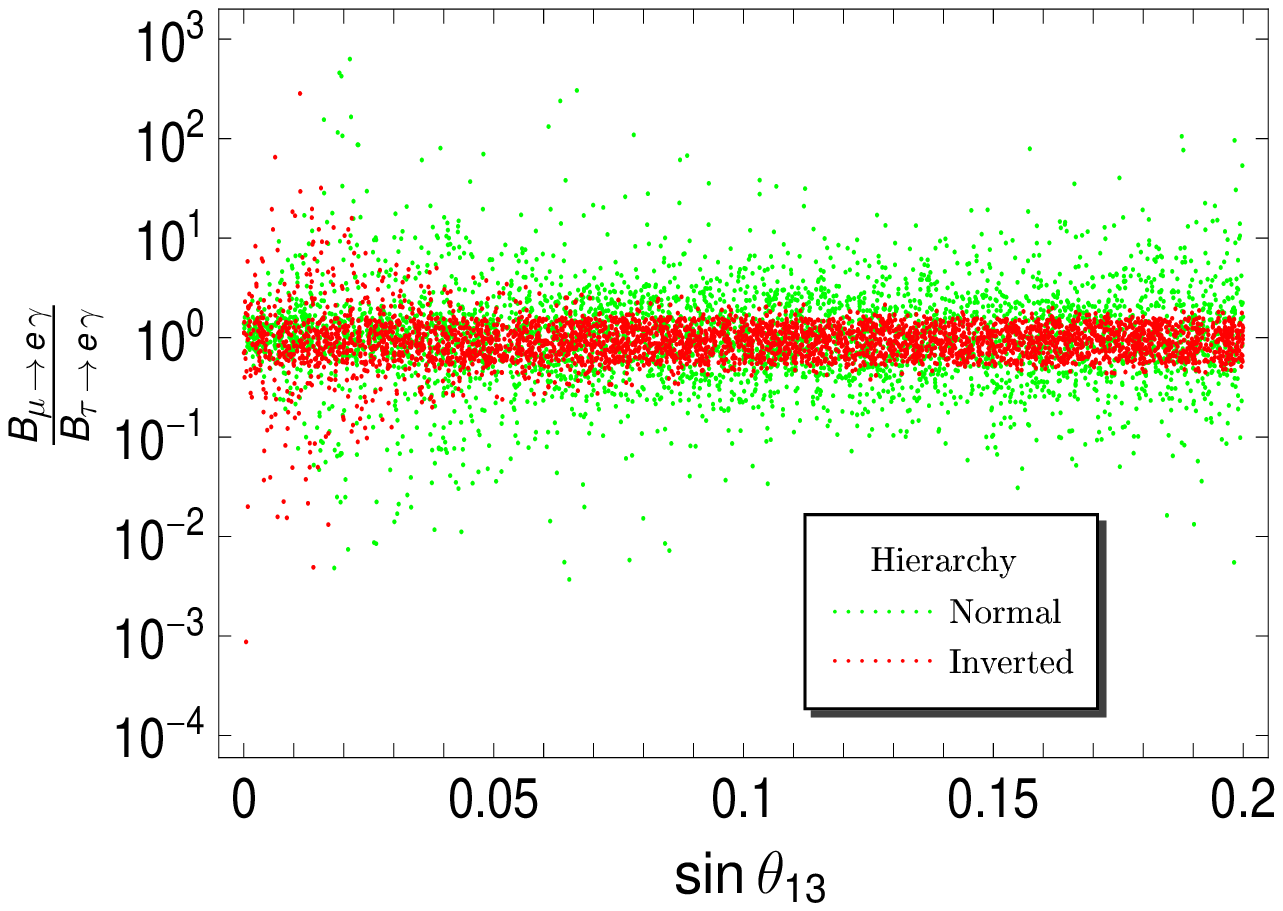}
  \caption{Same than Fig.~\ref{fig::brm1} but as a function of
    $\sin\theta_{13}$.}
\label{fig::brs13}  
\end{figure}

Results for the ratio of the $\mu$ and $\tau$  radiative decays into electrons 
are depicted in the right panel  in Fig.~\ref{fig::brm1}. 
For NH, in the $m_{\nu_1}\ll m_{\nu_{2,3}}$  limit   
and neglecting terms suppressed by $\theta_{13}$ we obtain 
\begin{equation}
  \label{eq:B2}
\frac{B_{\mu\,\rightarrow\,e\,\gamma}}{B_{\tau\,\rightarrow\,e\,\gamma}} 
\ \sim \ 
\cot_{23}^2\,,
\end{equation}
while in the IH and  quasi degenerate case of large masses  
\begin{equation}
  \label{eq:B3}
\frac{B_{\mu\,\rightarrow\,e\,\gamma}}{B_{\tau\,\rightarrow\,e\,\gamma}} 
\ \sim \ 
\tan_{23}^2\,
\end{equation}
is obtained, approximately independently of the values of $\delta$ and
$\theta_{13}$.  A glance at the second panel in Fig.~\ref{fig::brm1}
confirms in fact that the $\mu/\tau$ ratio for decays into electrons
remain centered around one for all values of $m_{\nu_l}$.  For NH,
$B_{\tau\,\rightarrow\,e\,\gamma}$ ($B_{\mu\,\rightarrow\,e\,\gamma}$)
can vanish when $\delta=0\,(\pi)$ and
$\frac{m^2_{\nu_1} +m^2_{\nu_2}}{2 m^2_{\nu_3}}\approx \frac{\Delta\,m^2_{\rm
    sol}}{\Delta\,m^2_{\rm atm}} \frac{s_{12}c_{12}}{s_{13}}$, which
can be satisfied with $m_{\nu_1}$ in the few$\times 10^{-2}\,$eV
range.  This accidental enhancement (suppression) of the ratio can be
clearly seen in the central region of the second scatter plot.

For completeness, we have also studied the dependence of these two
ratios as a function of $\sin\theta_{13}$ and of the Dirac phase
$\delta$. The first panel in Fig.~\ref{fig::brs13} confirms that for
NH a factor of a few enhancement of $B_{\mu\,\rightarrow\,e\,\gamma}$
with respect to $B_{\tau\,\rightarrow\,\mu\,\gamma}$ is possible,
independently of the value of $\theta_{13}$. For IH instead,
$B_{\mu\,\rightarrow\,e\,\gamma}$ is suppressed as $\theta_{13}^2\,$, in
agreement with~\eqn{eq:B1a}.
\begin{figure}[t!]
  \centering
  \includegraphics[width=7.5cm,height=6.0cm]{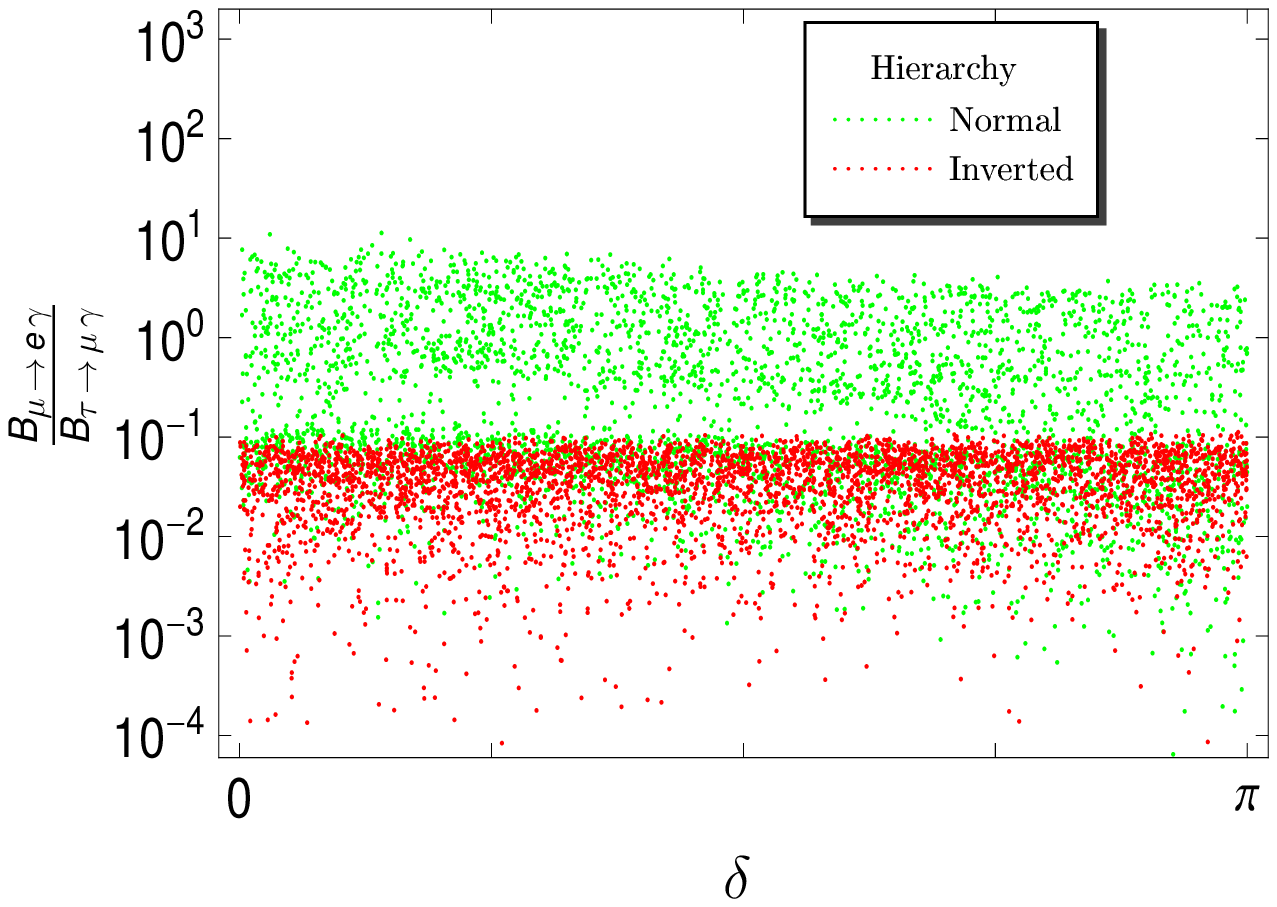}
  \includegraphics[width=7.5cm,height=6.0cm]{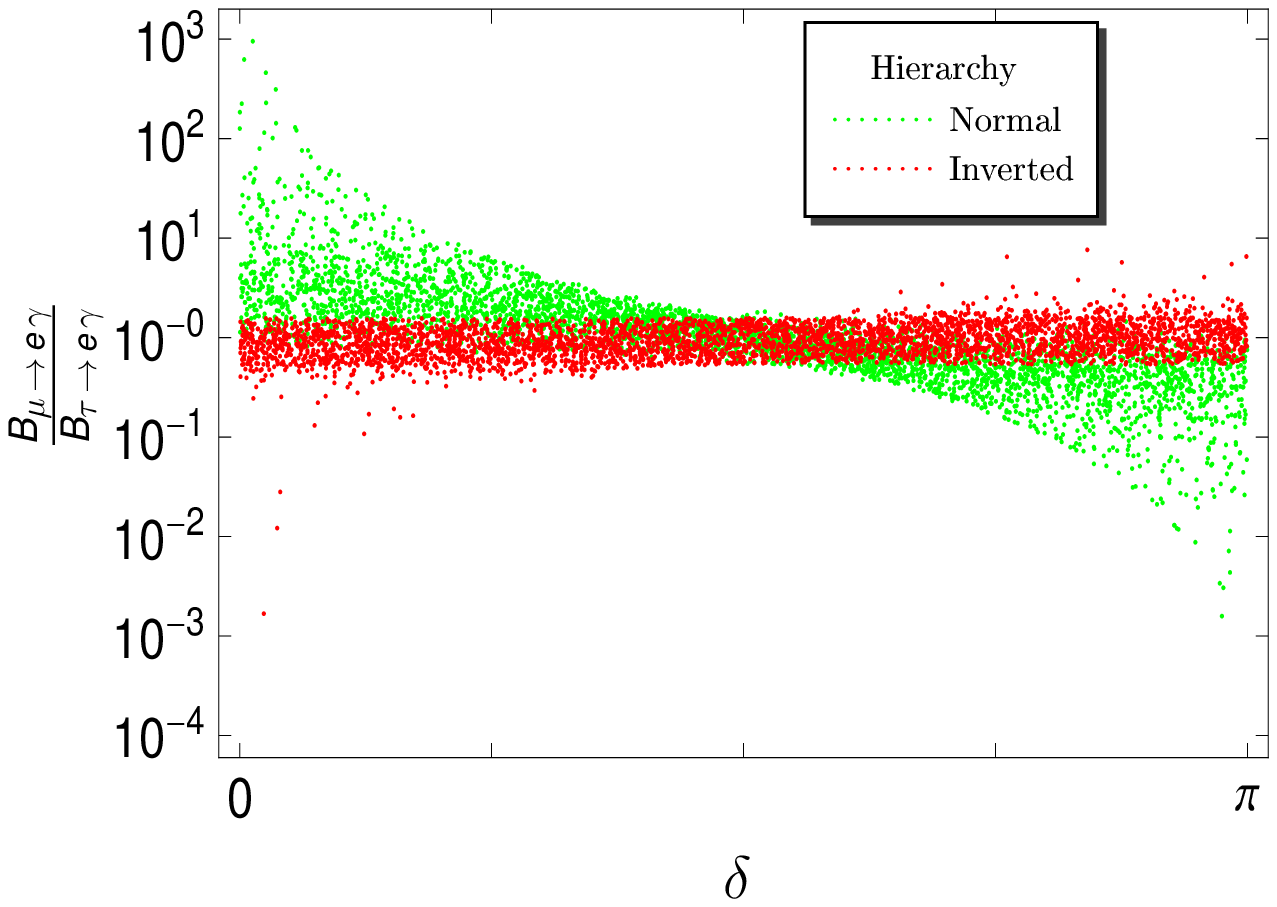}
\caption{Same than Fig.~\ref{fig::brm1} but as a function of 
the Dirac phase $\delta$.}
\label{fig::brdelta}  
\end{figure}

Fig.~\ref{fig::brdelta} depicts the correlations of the results with
the Dirac phase $\delta$. For NH the enhancements of
$B_{\mu\,\rightarrow\,e\,\gamma}$ can occur for all values of the
phase.  In contrast, $B_{\mu\,\rightarrow\,e\,\gamma}$
can dominate over $B_{\tau\,\rightarrow\,e\,\gamma}$ only 
if $\delta <\frac{\pi}{2}$.  Note that the extremely large enhancements
(suppressions) that can be seen in the second panel of this figure for
$\delta\to 0$ ($\delta\to \pi$) are not parametric effects, but arise
from the already mentioned accidental cancellations that can occur 
when $m_{\nu_1}\sim 10^{-2}\,$eV.

In summary, the new MLFV scenario that we have been discussing is
characterized by a quite different phenomenology from the case
previously studied in~\cite{Cirigliano:2005ck} since, in contrast to
that case, it allows the branching fraction
$B_{\mu\,\rightarrow\,e\,\gamma}$ to dominate over
$B_{\tau\,\rightarrow\,\mu\,\gamma}$ and
$B_{\tau\,\rightarrow\,e\,\gamma}$.  The enhancement with respect
$B_{\tau\,\rightarrow\,\mu\,\gamma}$ that occurs in the NH case does
not exceed a factor of a few, but it is parametric in the small values
of $m_{\nu_1}$.  The strong enhancement with respect to
$B_{\tau\,\rightarrow\,e\,\gamma}$ instead is due to accidental
cancellations that suppress this process, and that become particularly
efficient when $\delta$ is close to zero.

Needless to say, since the ratio of normalized branching ratios of
other LFV processes like for example $B_{\mu\to 3e}$, $B_{\tau\to
  3\mu}$, $B_{\tau\to 3e}$ are controlled by the same LFV factors
$\Delta^{(2)}_8$, they are  characterized by a completely similar
pattern of enhancements/suppressions.

In view of the ongoing high sensitivity experimental searches for
LFV processes~\cite{MEG} besides comparing the rates for different LFV
channels it is also of primary interest to give an estimate of the
absolute values of the branching fractions. In the most favorable
case, in which  $\Delta^{(2)}_8$ is a matrix with  $\cO(1)$  entries,  
it is easy to derive a rough estimate: 
\begin{equation}
  \label{eq:absolute}
 B_{\mu\,\rightarrow\,e\,\gamma}\ \approx\ 
1536\pi^3 \alpha \frac{v^4}{\Lambda^4}.
\end{equation}
Comparing~\eqn{eq:absolute} with the present experimental limit $
B^{\rm exp}_{\mu\,\rightarrow\,e\,\gamma}\lsim 10^{-11}$~\cite{MEGA}
we can conclude that the scale of LFV should be rather large: $\Lambda
\gsim 400\,$TeV.

\section{Conclusions}

In this paper we have studied MLFV extensions of the type-I seesaw.
This model is characterized by the group of broken symmetries
$SU(3)_\ell\times SU(3)_N\times SU(3)_e\times U(1)_L\times U(1)_R$,
where the last Abelian factor $U(1)_R$ can be identified with phase
rotations of the RH leptons $e$ or of the RH neutrinos $N$, being this
second choice phenomenologically more interesting and thus the one
that we have adopted.  We have parametrized the breaking of the
Abelian symmetries by means of two spurions $\mu_L$ and
$\epsilon_\nu$, and the breaking of the semi-simple flavour group
$SU(3)^3$ by means of three `non-Abelian' spurions $Y_e\,, Y_\nu$ and
$Y_M$, being the first two related to the charged leptons and RH
neutrinos Yukawa couplings and the last one with the RH neutrinos
mass matrix.

We have seen that formulating a predictive MLFV framework, that is a
framework in which LFV effects can be completely described in terms of
low energy observables, is possible only if the number of relevant LFV
free parameters of the seesaw is reduced. This can be achieved by imposing
specific conditions on the structure of the non-Abelian spurions.  We
have identified two basic possibilities that can be elegantly realized
by assuming from the beginning that the underlying symmetry of the
type-I seesaw  is a subgroup of the full flavour symmetry. These two
possibilities are:
\begin{itemize}
\item[{\bf A.}]  $Y_\nu$ is proportional to a unitary matrix. This  
  means that there is a basis in which $Y_\nu\sim I_{3\times 3}$ being
  $I_{3\times 3}$ the identity in flavour space.  This scenario can be
  realized by restricting the flavour symmetry $SU(3)_e\times SU(3)_N$
  to its subgroup $SU(3)_{\ell+N}$.
\item[{\bf B.}] $Y_M\sim I_{3\times 3}$ and $Y_\nu$ is a matrix with
  real entries $Y_\nu=Y_\nu^*$.  The symmetry reduction that realizes
  this second possibility is $SU(3)_N\to O(3)_N$. In this case CP
  conservation in the lepton sector must be also imposed to ensure the
  reality of $Y_\nu$.
\end{itemize}
While the second possibility is well
studied~\cite{Cirigliano:2005ck,Cirigliano:2006su}, the first one is
new, and yields a quite different phenomenology from case {\bf
  B}. Most remarkably, it allows for sizable enhancements of
processes involving the $\mu \to e$ transition with respect to
LFV processes involving the $\tau$ lepton.

As regards the two broken Abelian factors $U(1)_L$ and $U(1)_N$, we
have found that by parametrizing their breaking independently from the
breaking of the semi-simple groups of flavour transformations leaves
open unexpected possibilities: (i) the mass scale of the RH neutrinos
gets decoupled from the large seesaw scale $\mu_L$, since the former
breaks $U(1)_N$ by two units while, in the low energy effective
theory, the latter only breaks $U(1)_L$.  If $U(1)_N$ breaking is
small, say $\epsilon_\nu< 10^{-5}$, then the RH neutrinos can be at the TeV
scale or even below.  (ii) Higher dimension operators, like $(\bar N
N) (\bar q q)$, that could be effective for producing the RH neutrinos, 
do not break $U(1)_N$, and therefore are not affected by any 
strong suppression of the RH neutrino masses.  Of course, in our framework, 
such an intriguing scenario represents just  an open possibility that is 
allowed by the symmetries of the type-I seesaw. It remains to
be seen if explicit models realizing this scenario can be constructed.

\section*{Acknowledgments}
R.~A. acknowledges CICYT through the project FPA2009-09017, CAM
through the project HEPHACOS, P-ESP-00346, and financial support from
the MICINN grant BES-2010-037869.  G.~I.~acknowledges the support of
the Technische Universit\"at M\"unchen -- Institute for Advanced
Study, funded by the German Excellence Initiative.  
L.~M.~acknowledges the German `Bundesministerium f\"ur
Bildung und Forschung' under contract 05H09WOE.
R.~A.~and L.~M.~thank the theoretical group of the
Laboratori Nazionale di Frascati for hospitality
during the development of this project. 


   \end{document}